\documentclass[prl,twocolumn,floats]{revtex4}

\usepackage{graphicx}

\begin{document}

\title{Invariant Imbedding Equations for Electromagnetic
Waves in Stratified Magnetic Media: Applications to
One-Dimensional Photonic Crystals}

\author{Kihong \surname{Kim}}
\email{khkim@ajou.ac.kr}
\author{H. \surname{Lim}}
\affiliation{Department of
Molecular Science and Technology, Ajou University, Suwon 442-749,
Korea}
\author{Dong-Hun \surname{Lee}}
\affiliation{Department of Astronomy and Space Science, Kyung Hee
University, Kyunggi, Korea}

%\date{\today}

\begin{abstract}
We derive the invariant imbedding equations for plane electromagnetic
waves propagating in stratified magnetic media, where both
dielectric and magnetic permeabilities vary in one spatial direction
in an arbitrary manner. These equations allow us to
obtain the reflection and transmission coefficients of the waves and the
field amplitudes inside the media exactly
for any polarization and incident angle of the incoming wave by solving
an initial value problem of a small number of ordinary differential
equations. We apply our results to one-dimensional photonic crystals,
where the periodic variations of both dielectric and magnetic
permeabilities create photonic band gaps in the frequency spectrum.
\end{abstract}

\pacs{PACS Numbers: 42.25.Bs, 42.25.Dd, 42.70.Qs}

\maketitle

The propagation of electromagnetic waves in inhomogeneous
media is an important topic in various branches of physics,
including optics, plasma physics, astrophysics,
and condensed matter physics \cite{yeh,ginz,chandra,sheng}.
Great attention has been paid to the cases where the inhomogeneity
is random or periodic \cite{yeh,sheng,ishi,krav,kly1,kush,jkps1,jkps2}.
The problem of electromagnetic
wave propagation in random dielectric media
is analogous to the Anderson localization problem
of noninteracting electrons in a random potential \cite{sheng,john1,john2,crete}.
In recent years, much
effort has been made to demonstrate the localization of light experimentally
\cite{lagen,gena}
and to use the effect for developing the so-called random laser \cite{lawa,cao}.
Another important area of research has been created from an analogy
with the electronic band structure problem in condensed matter physics.
In media where the dielectric permeability varies
periodically in space, a series of photonic band gaps can appear in their
frequency spectra \cite{john2,yab1,yab2}.
Because of the promising possibility that a number of devices
based on this phenomenon can be used in various optoelectronic applications,
research activity in photonic band gap structures has increased explosively in
the last decade \cite{crete,mit}.

A number of analytical and numerical techniques have been developed
for analyzing the wave propagation
in inhomogeneous media \cite{yeh,ginz,chandra,sheng,ishi,krav,kly1,pendry,jkps3}.
In the most general cases where the inhomogeneity
is three-dimensional, the exact solution of the problem is extremely hard
to obtain, and one is forced to use some approximation scheme.
In the cases where the inhomogeneity is one-dimensional,
however, there exist several theoretical methods
that allow (at least numerically) exact
solutions of the problem in some simple situations.
The invariant imbedding method is a powerful and most
versatile tool for handling one-dimensional situations \cite{kly1,kly2,kly3,ram1,ram2,kim}.
It can be used to
obtain exact solutions for the reflection and transmission
coefficients of incoming waves and the electric and magnetic field
amplitudes inside arbitrarily
inhomogeneous media. When the inhomogeneity is random,
it can be used to obtain the exact disorder-averaged reflection and
transmission coefficients and field amplitudes \cite{kim}. The invariant imbedding method
can also be used to study the wave propagation in nonlinear dielectric
media \cite{kly2,ram2}
and the propagation of several coupled waves \cite{kly3} in an exact manner.

In the present letter, we apply the invariant imbedding method
to cases where both the dielectric and magnetic permeabilities
vary in one direction in space in an arbitrary manner and
derive a set of first-order
ordinary differential equations called the invariant
imbedding equations.
To the best of our knowledge, these equations have never been derived before.
We solve the initial value problem of
the invariant imbedding equations numerically
to obtain the exact reflection and transmission coefficients and field
amplitudes.
We apply our results to the cases
where one-dimensional periodic variations of both the dielectric and magnetic
permeabilities create photonic band gaps in the frequency spectrum.
This situation
can be realized in one-dimensional magnetic photonic crystals
made of alternating layers of dielectric and ferrite films \cite{sou,kee2,kee3,kee4}.

We are interested in the propagation of a plane, monochromatic, and
linearly-polarized electromagnetic wave of angular frequency
$\omega$ and vacuum wave number $k_0=\omega/c$, where $c$ is the
speed of light in vacuum. The wave is assumed to be incident
from a homogeneous region on a layered or
stratified medium, where the dielectric permeability $\epsilon$
and the magnetic permeability $\mu$ (and therefore the refractive
index $n=\sqrt{\epsilon\mu}$ and the wave impedance
$Z=\sqrt{\mu/\epsilon}$) vary only in one direction in space. We
take this direction as the $z$-axis and assume the inhomogeneous,
but isotropic, medium lies in $0 \le z \le L$. Without loss of
generality, we assume that the wave propagates in the $xz$-plane.
Since the medium is uniform in the $x$-direction, the dependence
on $x$ can be taken as being through a factor $e^{iqx}$, with $q$ being 
a constant. When $q=0$, the wave passes through the medium
normally. If $q\ne 0$, the wave is said to pass obliquely.

For $q\ne 0$, we have to distinguish two independent cases of polarization.
In the first case, the electric field vector
is perpendicular to the $xz$-plane.
Then one can show easily from Maxwell's equations that
the complex amplitude of the electric field, $E=E(z)$,
satisfies
\begin{equation}
{{d^2E}\over{dz^2}}-\frac{1}{\mu(z)}\frac{d\mu}{dz}
\frac{dE}{dz}+\left[k_0^2\epsilon(z)\mu(z)-q^2\right]E=0.
\label{eq:s}
\end{equation}
This type of wave is known as an $s$ (or TE) wave.
In the other case, the magnetic field vector is perpendicular to
the $xz$-plane. Then
the complex amplitude of the magnetic field, $H=H(z)$,
satisfies
\begin{equation}
{{d^2H}\over{dz^2}}-\frac{1}{\epsilon(z)}\frac{d\epsilon}{dz}
\frac{dH}{dz}+\left[k_0^2\epsilon(z)\mu(z)-q^2\right]H=0.
\label{eq:p}
\end{equation}
This type of wave is called $p$ (or TM) wave. We note that there
is a simple symmetry between Eqs.~(\ref{eq:s}) and (\ref{eq:p}):
Eq.~(\ref{eq:p}) is obtained from Eq.~(\ref{eq:s}) by exchanging
$\epsilon$ and $\mu$ and replacing $E$ by $H$.
From now on, we will consider only the $s$ wave. The results for the
$p$ wave follow from those for the $s$ wave in a trivial manner.

We assume that the wave is incident from the region where $z>L$
and is transmitted to the region where $z<0$. The dielectric and
magnetic permeabilities are assumed to be given by
\begin{eqnarray}
\epsilon(z)&=&\left\{\begin{array}{ll}
\epsilon_1 & ~~\mbox{if $z>L$}\\
\epsilon_R(z)+i\epsilon_I(z) & ~~\mbox{if $0 \le z \le L$}\\
\epsilon_2 & ~~\mbox{if $z<0$}\end{array}\right.,
\nonumber\\
\mu(z)&=&\left\{\begin{array}{ll}
\mu_1 & ~~\mbox{if $z>L$}\\
\mu_R(z)+i\mu_I(z) & ~~\mbox{if $0 \le z \le L$}\\
\mu_2 & ~~\mbox{if $z<0$}\end{array}\right.,
\end{eqnarray}
where $\epsilon_1$, $\epsilon_2$, $\mu_1$, and $\mu_2$ are real
constants and $\epsilon_R(z)$, $\epsilon_I(z)$, $\mu_R(z)$, and
$\mu_I(z)$ are {\it arbitrary} real functions of $z$. When
$\theta$ is defined as the angle of incidence, the
constant $q$ in Eqs.~(\ref{eq:s}) and (\ref{eq:p}) is equal to
$\sqrt{\epsilon_1\mu_1}k_0\sin\theta$.

We consider a plane wave of unit magnitude $\tilde
E(x,z)=E(z)e^{iqx}=e^{ip(L-z)+iqx}$, where
$p=\sqrt{\epsilon_1\mu_1}k_0\cos\theta$, incident on the medium
from the right. The quantities
of main interest are the complex reflection and transmission
coefficients, $r=r(L)$ and $t=t(L)$, defined by the wave
functions outside the medium:
\begin{equation}
\tilde E(x,z)=\left\{ \begin{array}{ll}
e^{ip(L-z)+iqx}+r(L)e^{ip(z-L)+iqx},  &  ~z>L \\
t(L)e^{-ip^{\prime}z+iqx},  &  ~z<0  \end{array} \right.,
\end{equation}
where $p^\prime =\sqrt{\epsilon_2\mu_2}k_0\cos\theta^{\prime}$ and
$\theta^\prime$ is the angle that outgoing waves make with the
negative $z$-axis.

We generalize the invariant imbedding method developed in Ref.~23
to the cases where both $\epsilon$ and $\mu$ are inhomogeneous.
Let us consider the $E$ field in Eq.~(\ref{eq:s}) as a function
of both $z$ and $L$: $E=E(z;L)$. Starting from the crucial
observation that the boundary value problem of
the wave equation, Eq.~(\ref{eq:s}), can be transformed into an integral
equation
\begin{eqnarray}
&&E(z;L)=g(z;L)+\frac{ip}{2}\int_0^L dz^\prime g\left(z;z^\prime\right)
\nonumber\\
&&~~~\times \left\{\frac{\epsilon(z^\prime)}{\epsilon_1}
-\frac{\mu(z^\prime)}{\mu_1}+\frac{q^2}{p^2}
\left[\frac{\epsilon(z^\prime)}{\epsilon_1}
-\frac{\mu_1}{\mu(z^\prime)}\right]\right\}
E(z^\prime;L),\nonumber\\
&&g(z;z^\prime)=\exp\left[ip~{\mathrm sgn}(z-z^\prime)
\int_{z^\prime}^{z}dz^{\prime\prime}~
\frac{\mu(z^{\prime\prime})}{\mu_1}\right], \label{eq:e}
\end{eqnarray}
we derive
\begin{equation}
\frac{\partial E(z;L)}{\partial L}=a(L)E(z;L), \label{eq:e2}
\end{equation}
where
\begin{eqnarray}
a(L)&=&ip\frac{\mu(L)}{\mu_1}
+\frac{ip}{2}\bigg\{
\frac{\epsilon(L)}{\epsilon_1}-\frac{\mu(L)}{\mu_1}\nonumber\\
&&+\frac{q^2}{p^2}\left[\frac{\epsilon(L)}{\epsilon_1}
-\frac{\mu_1}{\mu(L)}\right]\bigg\}E(L;L).\label{eq:b}
\end{eqnarray}
Using this equation, we obtain
{\it exact} differential equations satisfied by $r$ and $t$:
\begin{eqnarray}
&&{{dr(L)}\over{dL}}=2i\sqrt{\epsilon_1\mu_1}k_0\cos\theta~
\frac{\mu_R(L)+i\mu_I(L)}{\mu_1}r(L)
\nonumber\\
&&+{i\over 2}\sqrt{\epsilon_1\mu_1}k_0\cos\theta~
\bigg[\frac{\epsilon_R(L)+i\epsilon_I(L)}{\epsilon_1}
-\frac{\mu_R(L)+i\mu_I(L)}{\mu_1}\nonumber\\&&
~+\frac{\epsilon_R(L)+i\epsilon_I(L)}{\epsilon_1}\tan^2\theta
-\frac{\mu_1}{\mu_R(L)+i\mu_I(L)}\tan^2\theta\bigg]\nonumber\\&&
~~\times\left[1+r(L)\right]^2,\nonumber\\
&&{{dt(L)}\over{dL}}=i\sqrt{\epsilon_1\mu_1}k_0\cos\theta~
\frac{\mu_R(L)+i\mu_I(L)}{\mu_1}t(L)
\nonumber\\
&&+{i\over 2}\sqrt{\epsilon_1\mu_1}k_0\cos\theta~
\bigg[\frac{\epsilon_R(L)+i\epsilon_I(L)}{\epsilon_1}
-\frac{\mu_R(L)+i\mu_I(L)}{\mu_1}\nonumber\\&&
~+\frac{\epsilon_R(L)+i\epsilon_I(L)}{\epsilon_1}\tan^2\theta
-\frac{\mu_1}{\mu_R(L)+i\mu_I(L)}\tan^2\theta\bigg]\nonumber\\&&
~~\times \left[1+r(L)\right]t(L). \label{eq:rt}
\end{eqnarray}
These invariant imbedding equations are supplemented with the initial
conditions for $r$ and $t$, which are obtained using the
well-known Fresnel formulas:
\begin{eqnarray}
r(0)&=&\frac{\mu_2\sqrt{\epsilon_1\mu_1}\cos\theta
-\mu_1\sqrt{\epsilon_2\mu_2-\epsilon_1\mu_1\sin^2\theta}}
{\mu_2\sqrt{\epsilon_1\mu_1}\cos\theta
+\mu_1\sqrt{\epsilon_2\mu_2-\epsilon_1\mu_1\sin^2\theta}},\nonumber\\
t(0)&=&\frac{2\mu_2\sqrt{\epsilon_1\mu_1}\cos\theta}
{\mu_2\sqrt{\epsilon_1\mu_1}\cos\theta
+\mu_1\sqrt{\epsilon_2\mu_2-\epsilon_1\mu_1\sin^2\theta}}.
\label{eq:ic}
\end{eqnarray}

For given values of $\epsilon_1$, $\epsilon_2$, $\mu_1$, $\mu_2$,
$k_0$ (or $\omega$), and $\theta$ and for arbitrary functions
$\epsilon_R(L)$, $\epsilon_I(L)$, $\mu_R(L)$, and $\mu_I(L)$, we
solve the nonlinear ordinary differential equations in Eq.~(\ref{eq:rt})
numerically, using the initial conditions in Eq.~(\ref{eq:ic}), and
obtain the reflection and transmission coefficients $r$
and $t$ as functions of $L$. The
reflectivity $\mathcal{R}$ and the transmissivity
$\mathcal{T}$ are given by
\begin{equation}
{\mathcal R}=\vert r\vert^2,~~
{\mathcal T}=
\frac{\mu_1\sqrt{\epsilon_2\mu_2-\epsilon_1\mu_1\sin^2\theta}}
{\mu_2\sqrt{\epsilon_1\mu_1}\cos\theta}\vert t\vert^2.
\end{equation}
If the permeabilities have no imaginary parts, the quantities
$\mathcal{R}$ and $\mathcal{T}$ satisfy the law of
conservation of energy, ${\mathcal R}+{\mathcal T}=1$.

The invariant imbedding method can also be used in calculating
the field amplitude $E(z)$ inside the inhomogeneous medium.
We rewrite Eq.~(\ref{eq:e2}) as
\begin{eqnarray}
&&\frac{\partial E(z;l)}{\partial l}=i\sqrt{\epsilon_1\mu_1}k_0\cos\theta
\biggl\{
\frac{\mu_R(l)+i\mu_I(l)}{\mu_1}
\nonumber\\
&&~+\frac{1}{2}\bigg[\frac{\epsilon_R(l)+i\epsilon_I(l)}{\epsilon_1}
-\frac{\mu_R(l)+i\mu_I(l)}{\mu_1}\nonumber\\&&
~~+\frac{\epsilon_R(l)+i\epsilon_I(l)}{\epsilon_1}\tan^2\theta
-\frac{\mu_1}{\mu_R(l)+i\mu_I(l)}\tan^2\theta\bigg]\nonumber\\&&
~~\times \left[1+r(l)\right]\biggr\}
E(z;l).
\end{eqnarray}
For a given $z$ ($0<z<L$), the field amplitude $E(z;L)$
is obtained by integrating this equation from $l=z$ to $l=L$
using the initial condition $E(z;z)=1+r(z)$.

\begin{figure}
\includegraphics[width=.4\textwidth]{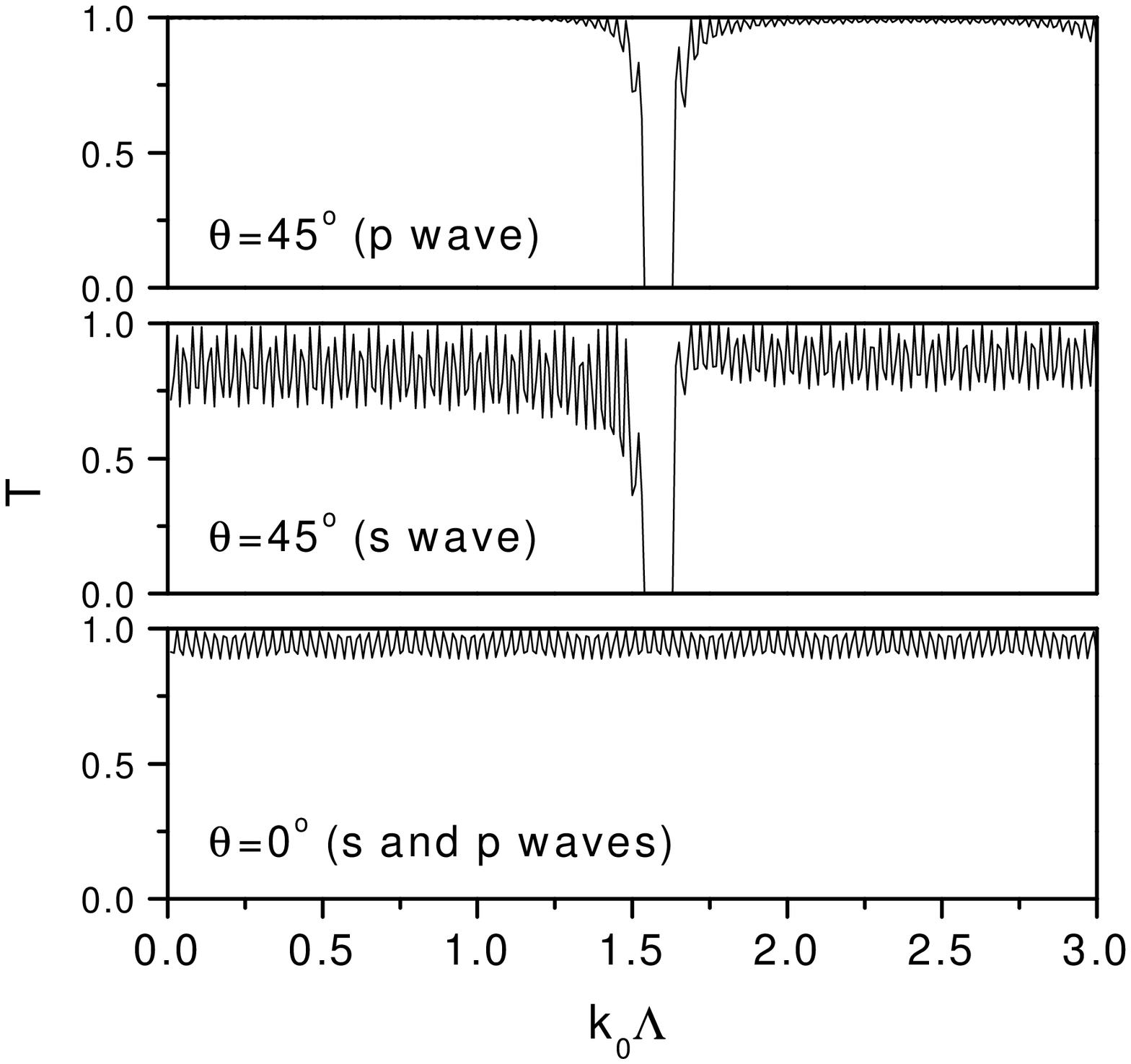}
\caption{Transmittance spectrum of a 100-period array
of alternating layers with $\epsilon_a=4$, $\mu_a=2$, $\epsilon_b=2$, $\mu_b=1$,
and $d_a=d_b=\Lambda/2$.}
\end{figure}

\begin{figure}
\includegraphics[width=.4\textwidth]{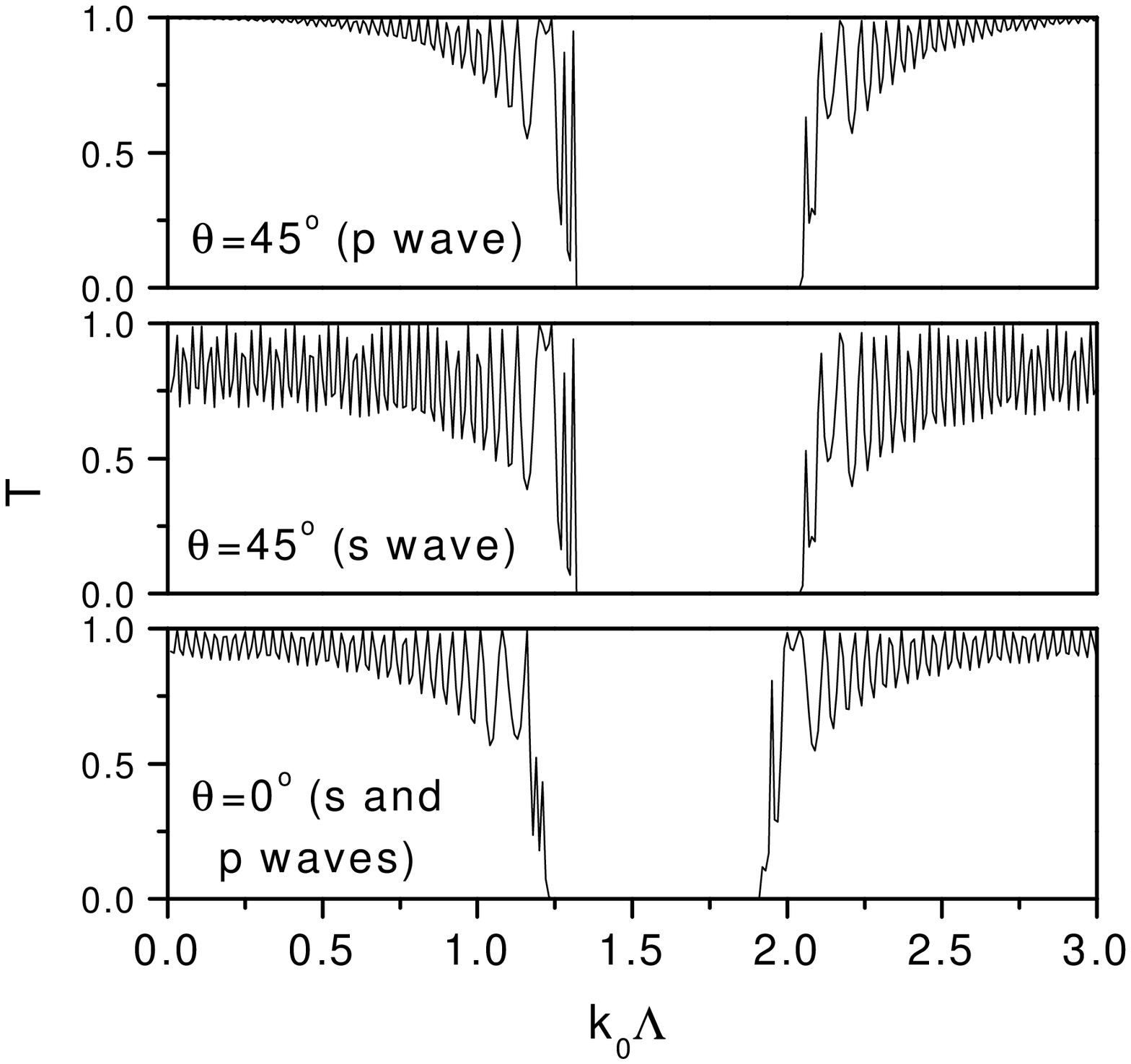}
\caption{Transmittance spectrum of a 100-period array
of alternating layers with $\epsilon_a=4$, $\mu_a=1$, $\epsilon_b=2$, $\mu_b=2$,
and $d_a=d_b=\Lambda/2$.}
\end{figure}

We demonstrate the validity and utility of our invariant imbedding equations
by applying them to simple one-dimensional photonic crystals,
where both $\epsilon$ and $\mu$ vary
periodically in one direction in space.
More specifically, we consider periodic arrays of alternating layers,
characterized by the dielectric permeabilities $\epsilon_a$ and $\epsilon_b$,
the magnetic permeabilities $\mu_a$ and $\mu_b$, and the thicknesses
$d_a$ and $d_b$. We assume that the imaginary parts of the
permeabilities are zero and the ambient medium is vacuum.
In the limit where the length of the system, $L$, diverges to infinity, one can find
the exact positions of the photonic band gaps 
by computing the roots of the analytical expression \cite{kee4}
\begin{equation}
\cos\beta_a\cos\beta_b-{1 \over 2}\left({{h_a} \over{h_b}}+{{h_b}\over{h_a}}\right)
\sin\beta_a\sin\beta_b +1=0,
\label{eq:root}
\end{equation}
where $n_i=\sqrt{\epsilon_i\mu_i}$, $Z_i=\sqrt{{\mu_i}/{\epsilon_i}}$ ($i=a,b$) and
\begin{eqnarray}
&&h_i=\left\{\begin{array}{ll}\sqrt{{n_i}^2-\sin^2\theta}/(n_iZ_i)& \mbox{for the $s$ wave}\\  
Z_i\sqrt{{n_i}^2-\sin^2\theta}/n_i& \mbox{for the $p$ wave}\end{array}\right.,\nonumber\\
&&\beta_i=k_0d_i\sqrt{{n_i}^2-\sin^2\theta}.    
\end{eqnarray}

In Fig. 1, we plot the transmittance $T$ ($\equiv\vert t\vert^2$) of $s$ and $p$ waves
impinging on
a 100-period array with $\epsilon_a=4$, $\mu_a=2$, $\epsilon_b=2$, $\mu_b=1$,
and $d_a=d_b$ as a function of $k_0 \Lambda$ where $\Lambda=d_a +d_b$. In this
case, the refractive index alternates periodically between
$n_a=2\sqrt{2}$ and $n_b=\sqrt{2}$, whereas the wave impedance is uniform.
We find that there is no photonic band gap when the wave is normally incident.
For obliquely incident waves, small gaps are created.
Furthermore, we observe that the band gaps for the $s$ wave agree precisely with those for
the $p$ wave in their sizes and positions.
When $\theta=45^\circ$ and $L\rightarrow \infty$, we find 
from Eq.~(\ref{eq:root}) that the lowest band gap exists
for $1.54<k_0\Lambda<1.63$, which is fully consistent with our
numerical result obtained using the invariant imbedding method.

In Fig. 2, we plot $T$ for the case of
a 100-period array with $\epsilon_a=4$, $\mu_a=1$, $\epsilon_b=2$, $\mu_b=2$,
and $d_a=d_b$. In this case, the wave impedance
alternates periodically between
$Z_a=1/2$ and $Z_b=1$, while the refractive index is uniform.
We find that large band gaps are formed for all incident angles. 
Similarly to the previous case, the band gaps for the $s$ wave agree with those for
the $p$ wave in their sizes and positions.
When $\theta=0^\circ$ and $45^\circ$, the first gaps are predicted to exist
for $1.23<k_0\Lambda<1.91$ and for $1.32<k_0\Lambda<2.04$, respectively,
in agreement with our result. 

From the two examples discussed above, we conclude that the wave impedance,
not the refractive index, 
is a crucial parameter mainly responsible for the creation of photonic
band gaps, as has been extensively discussed in Refs. 29 to 31.
The equality of the $s$ and $p$ wave gaps is a special feature that 
works only when either the refractive index or the wave impedance is uniform.
It is easy to verify this by examining Eq.~(\ref{eq:root}).

It is straightforward to apply our equations to more realistic cases where
the imaginary parts of the permeabilities are nonzero. They
can also be applied to cases where the parameters contain randomly-varying components.
This versatility of the invariant imbedding method makes it a very useful
and convenient tool in designing practical optical devices for various application
purposes.

\acknowledgments
This work has been supported by the Korea Research Foundation through Grant No.
KRF-2001-015-DP0167.

\end{document}